# Interpreting Paper-to-Patent Citations in AI-Related PMC Articles

Haowei Li[1] and Yi Bu[1,2]

*1. Department of Information Management, Peking University, Beijing 100871, China*

*2. Center for Informationalization and Information Management Research, Peking University, Beijing 100871, China*

**Correspondence concerning this article should be addressed to Yi Bu (Email: buyi@pku.edu.cn).**

## Abstract

Paper-to-patent citations connect scientific and technological documents, but their interpretability depends on how patents are used in the citing text. This study examines citation contexts and paper-level outcome associations in AI-related articles in PubMed Central (PMC). Screening 189,073 articles identified 746 with explicitly tagged patent references (0.395%), including 713 with 1,544 identifiable in-text patent-citation instances. Three large language models classified these instances by represented content, motivation, textual position, and evaluative stance. Regression models examined associations among these dimensions, and propensity-score-matched regressions compared patent-citing papers with XML-screened controls. Background-oriented citations were predominantly neutral and introductory, whereas adoption-oriented citations were more closely associated with methodological positions and supportive evaluations. These patterns suggest heuristic configurations of contextualization, operationalization, and problematization, although classification overlap and exclusion of rare categories limit their interpretation. In matched samples, the patent-citing indicator was associated with lower scholarly-citation measures and more citations received from patents. These associations are observational, not causal. Together, these findings show that paper-to-patent citations encode heterogeneous forms of engagement with patented technology and should be interpreted as selective bibliographic signals rather than direct indicators of technology-to-science knowledge transfer or research value.

## 1. Introduction

Scientific research and technological invention are interconnected forms of knowledge production. Citation links between scientific publications and patents have consequently been used to investigate relations between science and technology. Most research has examined the science-to-technology direction, tracing how scientific publications contribute to patented invention, how closely scientific and technological frontiers are linked, and whether high-quality science supports valuable technologies (Ahmadpoor & Jones, 2017; Gittelman & Kogut, 2003; Marx & Fuegi, 2020, 2022; Narin et al., 1997; Poege et al., 2019). This literature has substantially advanced the measurement of science–technology linkages, including through methods that identify scientific citations located in patent full text rather than only on patent front pages. Much less is known about the opposite citation direction, in which scientific papers cite patents. Early research identified such "reverse" citations as a distinct bibliographic relation between scientific and technological documents (Glänzel & Meyer, 2003), while subsequent large-scale work showed that papers cite patents only rarely (Hsiao & Torvik, 2019). This asymmetry raises a more fundamental question: what does a paper-to-patent citation actually represent?

A bibliographic link alone cannot establish that patented knowledge has entered the scientific research process. Patents disclose algorithms, models, devices, architectures, workflows, and application systems, but scientific papers may invoke such information for substantially different purposes. A patent may provide technical input to a study, but it may also supply historical or technological background, document an existing application, establish a benchmark, acknowledge related prior art, or identify a limitation. More generally, citation-based indicators provide selective and incomplete traces of knowledge relations rather than direct observations of knowledge transfer (Meyer, 2000; Roach & Cohen, 2013). Failing to distinguish these functions can lead to several inferential errors. A background citation may be interpreted as evidence that patented technology was operationally adopted; a paper containing several contextual patent citations may appear more technologically dependent than one that substantively uses a single patented method; and patent-citation counts or ratios may be interpreted as the intensity of technology-to-science knowledge transfer even when the cited patents serve primarily rhetorical or descriptive purposes. Such interpretations conflate the existence of a bibliographic link with the depth and form of technological engagement. The key issue is therefore not whether patent citation constitutes a major scientific practice, but what inferences can legitimately be drawn from different forms of paper-to-patent citation. Estimating how often these links occur and assessing what they signify are distinct tasks. Even an infrequent indicator requires a defensible interpretation when it is used to characterize science–technology relations. The contribution sought here is therefore to clarify the meanings and limits of observed links within a defined corpus, rather than to establish that patent citation is a widespread channel of technological input into science.

First, a gap remains in understanding how the citation contexts of paper-to-patent links are systematically characterized and interpreted. Existing research has already established several important descriptive regularities. Hsiao and Torvik (2020), for example, showed that paper-to-patent citations tend to occur near the beginning of scientific articles, are frequently mentioned only once, and often provide background information, although citations in methodological sections suggest that some patents may also inform research procedures. These findings provide an important descriptive baseline, but location and frequency alone cannot distinguish substantively different forms of engagement with patented technology. A citation in an introduction may provide historical background, establish a technological gap, document deployment, or criticize an existing technology; similarly, a citation in a methods section may indicate actual adoption, comparison, or merely description. Citation-context research more broadly has shown that references may establish credibility, characterize prior work, support claims, and position new contributions within scientific discourse (Gilbert, 1977; Small, 1978), while computational full-text studies distinguish functions such as background, use, comparison, and criticism (Boyack et al., 2018; Cohan et al., 2019; Teufel et al., 2006). The first gap, therefore, is not the absence of descriptive evidence on where patent citations occur, but the lack of an integrated framework for characterizing what technological content is represented, why the patent is cited, where it is positioned, and how the cited technology is evaluated.

Second, a gap remains in understanding how these citation dimensions are systematically related. Patents are heterogeneous technical documents (Meyer, 2000), and the motivation for citing a patent cannot be inferred from the type of technological content cited alone (Ding et al., 2014). The same patented algorithm may be introduced as background, adopted as a method, used as a benchmark, or criticized for a limitation; conversely, the same citation motivation may involve an algorithm, device, system, workflow, or application. Citation position adds structural information because citation functions vary systematically across article sections (Boyack et al., 2018; Cohan et al., 2019), while evaluative stance captures a related but distinct dimension of citation meaning (Abu-Jbara et al., 2013; Budi & Yaniasih, 2023). It therefore remains unclear whether represented content, citation motivation, textual position, and evaluative stance form systematic and recurring configurations in paper-to-patent citation.

Third, a gap remains in understanding what paper-to-patent citation signals at the paper level. Studies of science–technology linkages typically examine whether scientific papers are subsequently cited by patents as an indicator of technological visibility or relevance (Ahmadpoor & Jones, 2017; Marx & Fuegi, 2020; Poege et al., 2019), while research on academic patenting and university–industry engagement has examined relationships with publication productivity, quality, and research direction (Azoulay et al., 2009; Perkmann et al., 2013). Producing a patent and citing an existing patent are analytically different activities, however, and neither literature establishes what the presence or intensity of patent citation signals at the paper level.

Scholarly citations, subsequent patent citations, clinical-trial references, news mentions, disruption, and recorded retraction capture different aspects of recorded attention, knowledge-network position, and publication status (Funk & Owen-Smith, 2017; Park et al., 2023). Examining these outcomes provides descriptive comparisons of the associations for different patent-citation measures, e.g., the presence, number, and citation ratio of patents. These comparisons do not treat patent-citing papers as a uniformly more important class of research or the selected outcomes as interchangeable measures of research value.

AI-related articles in PMC provide a tractable empirical setting because their structured full text permits patent references to be linked to local citation contexts and article sections. The corpus includes computational methods and applications involving devices, systems, and workflows, allowing different reported uses of patent documents to be examined. Its coverage is nevertheless shaped by PMC and should not be taken to represent the full AI literature or to establish patterns unique to AI. Within this setting, the study examines citation contexts, associations among citation dimensions, and paper-level outcome profiles through the following research questions:
**RQ1:** In what contexts do AI-related PMC articles cite patent documents, in terms of represented technical content, citation motivation, textual position, and evaluative stance?
**RQ2:** How are citation motivation, cited patent content, textual position, and evaluative stance associated with one another in AI-related PMC articles?
**RQ3:** How is patent citation associated with downstream scholarly-citation, patent-citation, clinical-trial, news, disruption, and recorded-retraction outcomes among AI-related PMC articles matched on observed characteristics?

This study constructs a paper-to-patent citation dataset from the full text of AI-related articles in PubMed Central. At the citation level, three large language models independently classify patent-citation contexts by represented content, motivation, position, and stance, and multinomial logit models examine associations among these dimensions. At the paper level, patent-citing papers are compared with XML-screened non-patent-citing controls using hierarchical matching constraints, propensity score matching, and post-matching regression. Together, these analyses examine both the contextual meaning of individual patent citations and the paper-level signals associated with their presence and intensity.

## 2. Methodology

### 2.1 Data

This study conducted a large-scale corpus screening to construct a full-text patent-citation dataset for AI-related PMC articles using full-text records from PubMed Central (PMC). PMC articles are available in structured JATS XML, which provides article metadata, section information, full text, and typed reference records. These features enabled patent references to be identified from structured metadata and subsequently located within the citing text.

The initial corpus comprised PMC articles published between 1990 and 2025 whose titles, abstracts, or keywords contained AI-related terms, including “artificial intelligence,” “machine learning,” “deep learning,” “natural language processing,” and “computer vision.” This procedure identified 189,073 AI-related papers. A paper was classified as patent-referencing when its XML reference list contained at least one reference marked with publication-type="patent". In total, 746 papers contained patent references, yielding 1,333 paper–patent pairs. Parsing the full-text XML located 1,544 in-text patent-citation instances in 713 of these papers. The remaining 33 papers listed patents in their reference lists but contained no identifiable in-text mention. The observed proportion of papers with explicitly tagged patent references was 746/189,073 (0.395%); the proportion with identifiable in-text patent citations was 713/189,073 (0.377%). These are detection rates under the XML-based procedure, rather than estimates of all patent use or all patent citation in AI research.

To obtain a broader range of paper-level indicators, PMC records were matched with OpenAlex (Priem et al., 2022) and SciSciNet-v2 (Lin et al., 2023) to obtain publication metadata, topic classifications, citation indicators, reference counts, authorship and institutional information, funding variables, and cross-domain impact indicators. The final data support analyses at two levels. Citation-instance-level data are used to examine patent citation contexts and relationships among citation characteristics. Paper-level data are used to compare patent-citing and non-patent-citing papers. **Figure 1** summarizes the complete workflow.

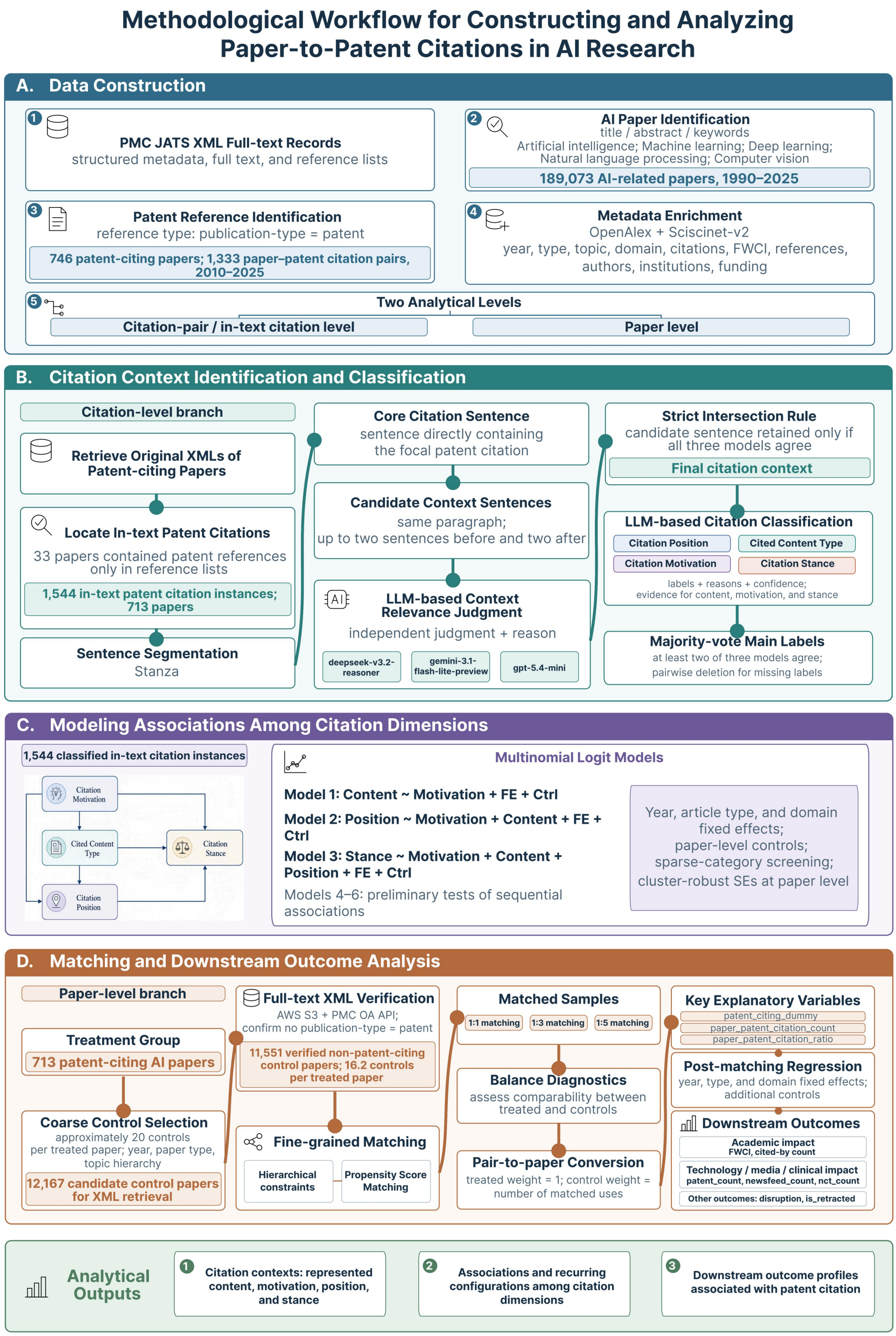

**Figure 1. Methodological Workflow for Constructing and Analyzing Paper-Patent Citations in AI-Related PMC Articles**

## 2.2 Methods

### 2.2.1 Citation Context Identification and Classification

The citation-context analysis began with the 1,544 in-text patent citation instances. Full texts were segmented into sentences using Stanza (Qi et al., 2020). For each instance, the sentence containing the focal patent citation was defined as the core citation sentence. Up to two preceding and two following sentences from the same paragraph were treated as candidate context sentences.

Three large language models, DeepSeek-V3.2-Reasoner, Gemini-3.1-Flash-Lite-Preview, and GPT-5.4-mini, independently evaluated whether each candidate sentence referred to the same patent or patented technological content as the core sentence. To construct a conservative context, a candidate sentence was retained only when all three models judged it relevant. When no candidate sentence received unanimous agreement, the context consisted only of the core sentence. The complete prompts, output schemas, and multi-citation disambiguation rules used for citation-context identification and classification are provided in **Supplementary Appendix S1**.

Each final context was classified along four dimensions: citation position, represented content type, citation motivation, and citation stance. **Table 1** summarizes the dimensions and categories; **Supplementary Table S1** provides complete operational definitions and category-specific decision rules.

**Table 1. Summary of Patent Citation Classification Dimensions**

| Dimension | Categories |
|---|---|
| Citation position | Eleven micro-level categories and five macro-level IMRaD categories |
| Cited content type | Algorithm/model, device/hardware, system architecture, application scenario, workflow/pipeline, resource/platform, unspecified |
| Citation motivation | Background, use/adoption, comparison/benchmark, motivation/gap, extension/improvement, critique/limitation, application/deployment |
| Citation stance | Supportive/adoptive, neutral/descriptive, critical/limiting, mixed |

Citation motivation was treated as a multi-label variable because one citation may serve several purposes simultaneously. Position, content type, and stance were treated as single-label variables. The three language models independently assigned labels, supplied confidence scores, and identified supporting textual evidence. For single-label dimensions, the final label was determined by majority agreement. For motivation, a label was retained when at least two models selected it. Cases in which all three models assigned different single labels were treated as missing.

Missingness was handled through pairwise deletion so that observations were excluded only from analyses requiring the missing dimension. Missing rates were 4.40% for micro-level position, 1.49% for macro-level position, 9.52% for content type, 0.52% for motivation, and 0.84% for stance.

### 2.2.2 Associations Among Patent-Citation Dimensions

The association analysis examined how motivation, content, position, and stance were related across the 1,544 citation instances. Citation motivation was treated as the analytical starting point because citation-function research defines intent in terms of why an author invokes a source and how prior knowledge is incorporated into the citing argument (Gilbert, 1977; Jurgens et al., 2018; Teufel et al., 2006). Background, adoption, comparison, and criticism therefore represent different rhetorical uses of patent documents (Cohan et al., 2019; Teufel et al., 2006).

Represented content type identifies the technological information that the citing paper attributes to the patent. This distinction is consistent with the view that cited documents may operate as representations of particular concepts, methods, or contributions rather than as undifferentiated documents (Jurgens et al., 2018; Small, 1978). Represented content type is consequently expected to be associated with motivation: application-oriented content may provide technological background or document deployment, whereas algorithms and workflows may be more relevant to adoption or comparison.

Citation position identifies where the patent enters the structure of the paper. Full-text studies show that citation locations vary systematically across article sections and that structural information is informative for identifying citation intent (Boyack et al., 2018; Cohan et al., 2019). Motivation and content are therefore expected to be associated with whether a patent citation appears in an introduction, methods section, results section, discussion, or another textual location.

Citation stance captures the evaluative relationship constructed toward the patented technology. Prior research distinguishes evaluative polarity from citation purpose while showing that the two can be analyzed jointly (Abu-Jbara et al., 2013). Citation function, role, and sentiment have likewise been treated as related dimensions of citation meaning (Budi & Yaniasih, 2023). Stance is therefore modeled as an evaluative outcome associated with motivation, cited content, and citation position. For example, adoption-oriented citations may be more supportive, whereas citations used to establish gaps or limitations may be more critical.

These propositions motivate a descriptive analytical ordering from rhetorical purpose to represented technical content, structural placement, and evaluative positioning. **Figure 1C** summarizes the proposed relationships. This ordering organizes the association models but does not identify a temporal or causal process.

Content type, position, and stance are nominal variables. The analysis used multinomial logit models, which reduce to binary logit when only two outcome categories remain after screening (McFadden, 1972). For outcome $Y_i$, category $j$, and reference category $j_0$, the general specification is:

$$\log\left[\frac{\Pr(Y_i = j)}{\Pr(Y_i = j_0)}\right] = \alpha_j + \boldsymbol{\beta}_j^T \boldsymbol{M}_i + \boldsymbol{\theta}_j^T \boldsymbol{C}_i + \boldsymbol{\phi}_j^T \boldsymbol{P}_i + \boldsymbol{\gamma}_j^T \boldsymbol{X}_i + \delta_{j,year(i)} + \lambda_{j,type(i)} + \mu_{j,domain(i)}, j \neq j_0 \quad (1)$$

where $\boldsymbol{M}_i$, $\boldsymbol{C}_i$, and $\boldsymbol{P}_i$ denote citation-motivation, content-type, and position indicators, respectively; $\boldsymbol{X}_i$ denotes control variables; and the remaining terms represent year, article-type, and domain fixed effects. Predictors not required for a particular model are omitted. The three main models are summarized in **Table 2**.

**Table 2. Outcome and Main Explanatory Variables in the Three Main Models**

| Model | Outcome | Main explanatory variables |
|---|---|---|
| Model 1 | Cited content type | Citation motivation |
| Model 2 | Citation position | Citation motivation and cited content type |
| Model 3 | Citation stance | Citation motivation, cited content type, and citation position |

All models controlled for country count, institution count, FWCI, citation count, reference count, patent-citation ratio, and the mean classification-confidence score for the dependent dimension. Rare outcome categories and predictor–outcome combinations were screened before estimation to reduce sparse-cell and quasi-separation problems.

Three reduced specifications were additionally estimated for descriptive coefficient comparisons rather than formal causal mediation tests; their equations and full estimates are reported in **Supplementary Appendix S3** and **Tables S2–S4**.

### 2.2.3 Matching and Downstream Outcome Analysis of Patent-Citing and Non-Patent-Citing Papers

The paper-level analysis examined whether patent citation was associated with subsequent scholarly, technological, clinical, media, disruption, and recorded-retraction outcomes. Because patent-citing papers may differ systematically from other papers, a two-stage procedure was used to construct a comparable control group. Potential controls were first selected according to publication year, article type, and the OpenAlex topic hierarchy, and their full-text XML reference lists were inspected to screen out records marked as containing patent references. The remaining controls were then matched to patent-citing papers by combining hierarchical constraints with propensity score matching (Rosenbaum & Rubin, 1983). The matching procedure prioritized agreement on publication year, article type, and topic, while allowing progressively broader disciplinary matches when more specific controls were unavailable. **Supplementary Appendix S4** and **Tables S5 and S7** report full details of control screening, matching constraints, propensity-score estimation, and balance diagnostics (Austin, 2011; Ho et al., 2007; Stuart, 2010).

Three matched samples were constructed using 1:1, 1:3, and 1:5 matching ratios, followed by weighted post-matching regressions. Three paper-level patent-citation measures were entered in separate models: a binary indicator, the number of patent citations, and the proportion of citations that were patents. The outcomes captured scholarly-citation measures, citations from subsequent patents, clinical-trial references, news mentions, disruption (Funk & Owen-Smith, 2017), and recorded retraction status. The regressions included publication year, paper type, and domain fixed effects, together with bibliographic, authorship, institutional, geographic, and funding covariates. **Supplementary Appendix S4** and **Table S6** provide detailed variable definitions, transformations, weighting procedures, and model specifications. Because matching and regression address only observed differences, the results are interpreted as adjusted associations rather than causal effects.

# 3. Results

## 3.1 Multidimensional Citation Contexts of Patent Citations

**Figure 2** summarizes the semantic characteristics of patent citations in the AI-related PMC sample. Algorithms or models and devices or hardware were the most frequently identified content types, whereas resources or platforms were uncommon. Background was the dominant citation motivation, while explicit extension or improvement was rare. Most citations were neutral and descriptive; supportive citations were more frequent than critical citations, and mixed evaluations were uncommon. At the structural level, citations were concentrated in introductory and prior-work sections, with more than half located within the macro-level Introduction. Identified patent-citation instances were concentrated after 2018 and occurred most often in research articles and Physical Sciences papers. **Supplementary Appendix S2** and **Figure S1** report additional descriptive distributions by publication year, article type, and domain.

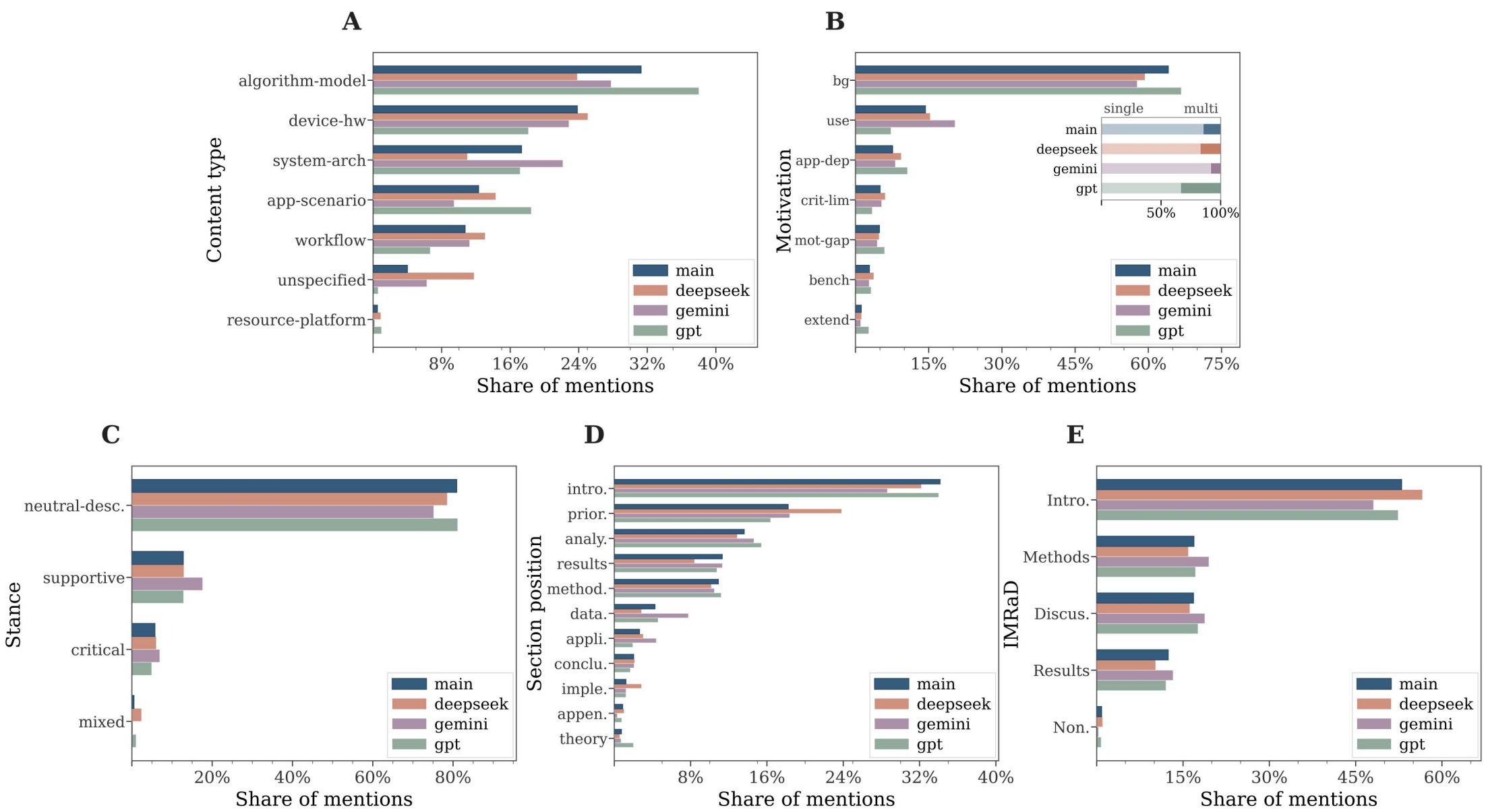

**Figure 2. Descriptive Patterns of Citation Contexts in AI-Related PMC Articles.** Panels A–E present the distributions of cited content, motivation, stance, and micro- and macro-level position. "Main" denotes the majority-vote labels, and the other bars show the labels from each individual model. Motivation is multi-label, so its category shares need not sum to 100%.

**Figure 3** reports the bivariate relationships among cited content, motivation, stance, and position. The heatmaps are row-normalized, and significant deviations from independence are based on false-discovery-rate-adjusted residual tests. Three main patterns emerged. First, cited content was associated with citation purpose. Application scenarios were disproportionately associated with background and deployment, whereas algorithms, models, and workflows were more frequently associated with adoption, comparison, or extension. Device- and hardware-related content appeared relatively more often in limitation-oriented contexts.

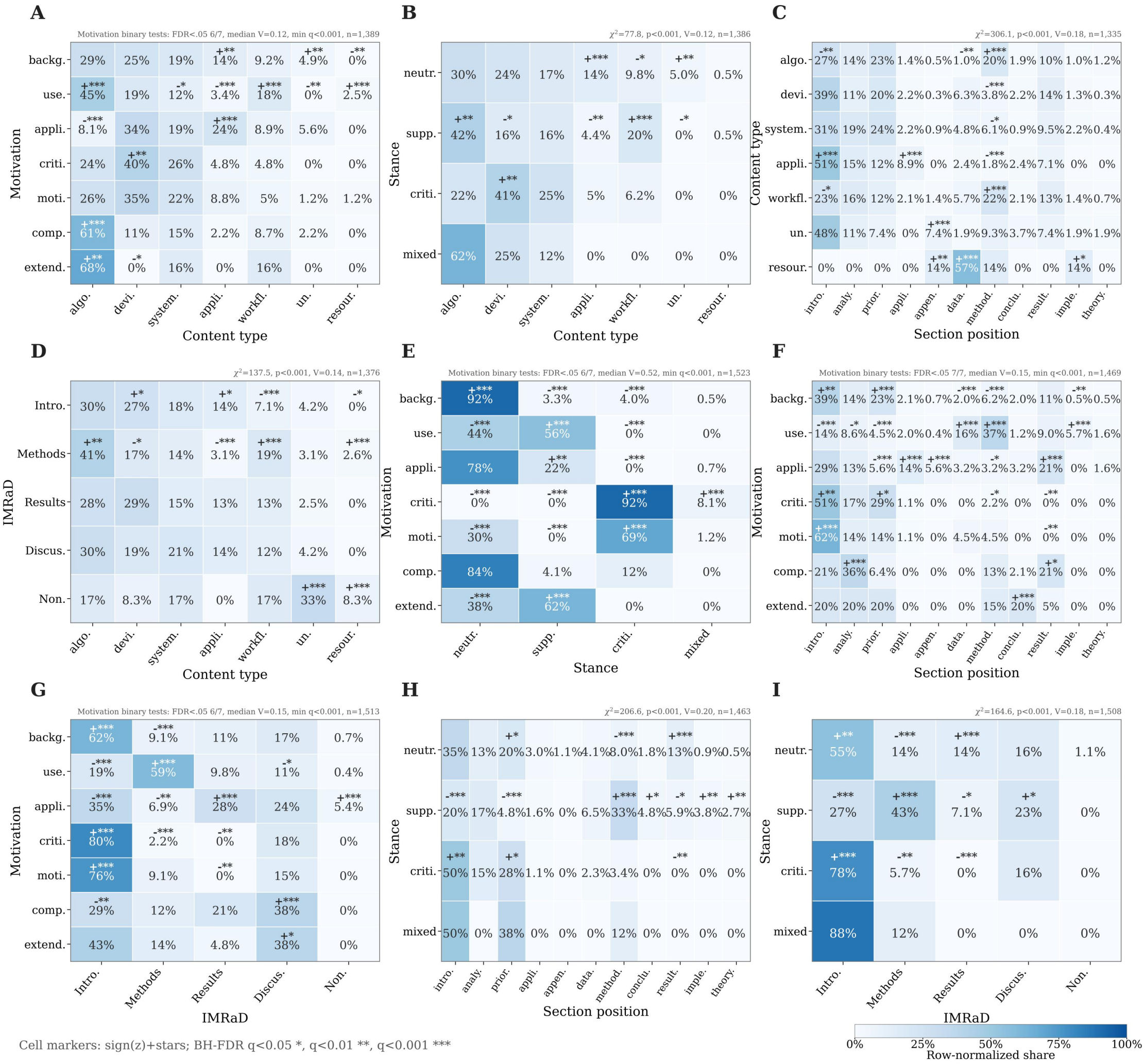

**Figure 3. Internal Associations Among Patent Citation Content, Motivation, Stance, and Position.** Panels A–D relate cited content to motivation, stance, and position. Panels E–G relate motivation to stance and position. Panels H–I relate stance to position. "+" and "–" indicate cells occurring significantly more or less often than expected under independence; stars indicate Benjamini–Hochberg false-discovery-rate-adjusted q-value thresholds.
* $q < .05$, ** $q < .01$, *** $q < .001$.

Second, motivation and content were associated with stance. Among instances with nonmissing motivation and stance labels, Figure 3E shows that 92% of background-labeled citations were neutral and 56% of use/adoption-labeled citations were supportive. These row-normalized percentages describe overlapping motivation groups, rather than mutually exclusive citation classes. Use/adoption-labeled citations were more frequently supportive than expected under independence. Gap-making and limitation-oriented citations were more frequently critical. Algorithms, models, and workflows were also more often associated with supportive evaluations, while application scenarios were predominantly descriptive and devices or hardware showed a comparatively stronger association with critical evaluation.

Third, the four dimensions were related to article structure. Background and gap-making citations were concentrated in introductory or prior-work sections. Adoption-oriented citations appeared more frequently in methods and implementation sections, while comparison, deployment, and improvement citations were relatively more common in results, application, or discussion sections. Algorithms, workflows, and platforms were concentrated in methods-related locations; application scenarios and devices appeared more often in introductory or application-oriented sections. Supportive citations occurred relatively more often in methods and discussion sections, whereas critical citations were more concentrated in introductions and prior-work reviews.

Additional heterogeneity analyses showed that patent-citation patterns varied by publication year, article type, and domain, especially in the distributions of represented content and citation position. **Supplementary Appendix S2** and **Figure S2** report the full results.

## 3.2 Conditional Associations Among Citation Motivation, Content, Position, and Stance

**Tables 3–5** report the three main multinomial logit models. **Supplementary Appendix S3** and **Tables S2–S4** report reduced specifications used to examine how estimated associations changed after represented content type or citation position was introduced. The reported average marginal effects estimate changes in category probabilities conditional on the included controls and fixed effects.

Model 1 showed that the presence of a background label, relative to its absence, was associated with the represented technological content. Conditional on the reported covariates, background motivation was associated with a 0.111 lower predicted probability of algorithm/model content and a 0.048 lower probability of workflow/pipeline content. It was associated with a 0.055 higher probability of application-scenario content and a 0.071 higher probability of system-architecture content. Its association with device/hardware content was not statistically significant. These estimates apply to the five retained content categories and do not compare background separately with each alternative motivation.

Model 2 showed differences in macro-level textual position by citation motivation. Background motivation was associated with a 0.204 higher probability of placement in the Introduction. Adoption motivation was associated with a 0.301 lower probability of introductory placement and a 0.397 higher probability of placement in Methods. Sparse-cell screening prevented estimation of the corresponding micro-level motivation effects. At the micro level, application scenarios, devices/hardware, and system architectures had higher probabilities of intro_motivation and lower probabilities of method_model_algorithm than algorithms/models. At the macro level, fewer content contrasts were statistically significant: application scenarios were less likely to appear in Methods and more likely to appear in Results; devices/hardware were more likely to appear in Introduction; and workflows/pipelines were less likely to appear in Introduction and more likely to appear in Methods. The reported macro-level contrasts for system architecture were not statistically significant.

Model 3 examined the two retained stance categories, neutral/descriptive and supportive/adoptive. Background motivation was associated with a higher probability of neutral description and a lower probability of supportive stance, whereas adoption motivation was associated with a 0.219–0.270 higher probability of supportive stance. Application/deployment motivation was positively associated with supportive stance in the macro-position specification. Application scenarios were less likely than algorithms/models to receive supportive evaluations in both specifications; the corresponding device/hardware contrast was statistically significant only in the macro-position specification. In that specification, Introduction, Methods, and Results were less likely to contain supportive evaluations than Discussion. In the micro-position specification, prior-work citations were less supportive than citations in analysis/discussion. These estimates do not cover critical or mixed stances and may partly reflect overlap between the motivation and stance definitions.

The reduced specifications showed that some coefficients changed after represented content type or citation position was added. **Supplementary Appendix S3** and **Tables S2–S4** report these comparisons. Because the labels were measured from the same citation contexts and the specifications are cross-sectional, coefficient attenuation is interpreted descriptively and does not identify mediation or an indirect causal pathway.

**Table 3. Multinomial Logit Estimates for Citation Motivation and Cited Content Type**

| Variables | | Model 1 | | | | |
|---|---|---|---|---|---|---|
| | | Cited Content Type | | | | |
| | | algorithm_model | application_scenario | device_hardware | system_architecture | workflow_pipeline |
| Citation Motivation | background | −0.111** | 0.055* | 0.033 | 0.071* | −0.048* |
| | | (0.035) | (0.022) | (0.030) | (0.028) | (0.022) |
| Controls | | Yes | | | | |
| Year Fixed Effects | | Yes | | | | |
| Article-Type Fixed Effects | | Yes | | | | |
| Domain Fixed Effects | | Yes | | | | |
| Observations | | 1,273 | | | | |
| Log Pseudolikelihood | | −1817.23 | | | | |

Notes: Average marginal effects (AMEs) from multinomial logit models are reported. Robust standard errors clustered at the paper level are shown in parentheses (Cameron & Miller, 2015). Retained motivation labels are entered as separate binary indicators because a patent citation may contain multiple motivation labels; their AMEs represent the average change in predicted probabilities when a given label is present rather than absent, holding other motivation labels and covariates constant. To reduce estimation instability caused by sparse cells and quasi-complete separation, outcome categories and cross-tabulated cells were pre-screened before model estimation. Categories with fewer than 100 observations or less than 2% of the estimation sample were excluded, rare fixed-effect categories were merged, and motivation indicators were retained only when the minimum cross-tabulated cell frequency was 20. As a result, the categories or labels included and reported in the table may be fewer than those in the full theoretical classification scheme. The predicted probabilities are conditional on membership in the retained outcome categories.

* p < .05, ** p < .01, *** p < .001.

**Table 4. Multinomial Logit Estimates for Citation Motivation, Cited Content Type, and Citation Position**

| | Model 2 | |
|---|---|---|
| Variables | Citation Position (micro) | Citation Position (macro) |

| | | analysis_discussion | intro_motivation | method_model_algorithm | prior_work_review | results_findings | Discussion | Introduction | Methods | Results |
|---|---|---|---|---|---|---|---|---|---|---|
| Citation Motivation | background | | | | | | −0.102** | 0.204*** | 0.006 | −0.108*** |
| | | | | | | | (0.039) | (0.048) | (0.031) | (0.030) |
| | use_adopt | | | | | | −0.063 | −0.301*** | 0.397*** | −0.032 |
| | | | | | | | (0.037) | (0.062) | (0.063) | (0.019) |
| Cited Content Type | application_scenario | 0.029 | 0.245*** | −0.176*** | −0.133** | 0.035 | 0.011 | 0.010 | −0.107*** | 0.086** |
| | | (0.035) | (0.046) | (0.025) | (0.042) | (0.027) | (0.038) | (0.042) | (0.029) | (0.033) |
| | device_hardware | −0.024 | 0.213*** | −0.150*** | −0.027 | −0.013 | −0.045 | 0.077* | −0.036 | 0.004 |
| | | (0.030) | (0.042) | (0.025) | (0.043) | (0.018) | (0.030) | (0.036) | (0.025) | (0.017) |
| | system_architecture | 0.012 | 0.121** | −0.124*** | 0.034 | −0.043** | −0.005 | 0.040 | −0.011 | −0.024 |
| | | (0.045) | (0.044) | (0.030) | (0.050) | (0.014) | (0.042) | (0.042) | (0.030) | (0.016) |
| | workflow_pipeline | 0.029 | −0.010 | 0.096 | −0.130*** | 0.016 | 0.006 | −0.124** | 0.103** | 0.014 |
| | | (0.042) | (0.048) | (0.052) | (0.039) | (0.028) | (0.037) | (0.045) | (0.035) | (0.026) |
| Controls | | Yes | | | | | Yes | | | |
| Year Fixed Effects | | Yes | | | | | Yes | | | |
| Article-Type Fixed Effects | | Yes | | | | | Yes | | | |
| Domain Fixed Effects | | Yes | | | | | Yes | | | |
| Observations | | 1,091 | | | | | 1,247 | | | |
| Log Pseudolikelihood | | −1113.8187 | | | | | −944.53742 | | | |

Notes: The reference category for Represented Content Type is algorithm_model. Blank micro-level motivation estimates indicate that the relevant motivation–outcome combinations did not satisfy the pre-estimation cell-frequency requirements. Other settings are the same as in **Table 3**.

* p < .05, ** p < .01, *** p < .001.

**Table 5. Multinomial Logit Estimates for Citation Motivation, Cited Content Type, Citation Position, and Citation Stance**

| Variables | | Model 3 | | | |
|---|---|---|---|---|---|
| | | Citation Stance | | | |
| | | neutral_descriptive | supportive_adoptive | neutral_descriptive | supportive_adoptive |
| Citation Motivation | background | 0.197** | −0.197** | 0.144** | −0.144** |
| | | (0.061) | (0.061) | (0.048) | (0.048) |
| | use_adopt | −0.219** | 0.219** | −0.270*** | 0.270*** |
| | | (0.073) | (0.073) | (0.073) | (0.073) |
| | application_deployment | | | −0.135*** | 0.135*** |
| | | | | (0.039) | (0.039) |
| Cited Content Type | application_scenario | 0.064* | −0.064* | 0.084** | −0.084** |
| | | (0.028) | (0.028) | (0.028) | (0.028) |
| | device_hardware | 0.004 | −0.004 | 0.053* | −0.053* |
| | | (0.024) | (0.024) | (0.022) | (0.022) |
| | system_architecture | −0.046 | 0.046 | −0.017 | 0.017 |
| | | (0.036) | (0.036) | (0.029) | (0.029) |
| | workflow_pipeline | −0.016 | 0.016 | −0.005 | 0.005 |
| | | (0.027) | (0.027) | (0.028) | (0.028) |
| Citation Position (micro) | intro_motivation | 0.028 | −0.028 | | |
| | | (0.032) | (0.032) | | |
| | method_model_algorithm | 0.019 | −0.019 | | |
| | | (0.035) | (0.035) | | |
| | prior_work_review | 0.086* | −0.086* | | |
| | | (0.035) | (0.035) | | |
| | results_findings | 0.075 | −0.075 | | |
| | | (0.044) | (0.044) | | |
| Citation Position (macro) | Introduction | | | 0.066* | −0.066* |
| | | | | (0.028) | (0.028) |
| | Methods | | | 0.068* | −0.068* |
| | | | | (0.029) | (0.029) |
| | Results | | | 0.099* | −0.099* |
| | | | | (0.041) | (0.041) |
| Controls | | Yes | | Yes | |
| Year Fixed Effects | | Yes | | Yes | |
| Article-Type Fixed Effects | | Yes | | Yes | |
| Domain Fixed Effects | | Yes | | Yes | |
| Observations | | 996 | | 1,155 | |
| Log Pseudolikelihood | | −208.94181 | | −270.72644 | |

Notes: The reference category for Represented Content Type is algorithm/model. The reference category for Citation Position (micro) is analysis/discussion, and the reference category for Citation Position (macro) is Discussion. The stance analysis includes only neutral_descriptive and supportive_adoptive because the other stance categories did not satisfy the pre-estimation frequency requirements. Other settings are the same as in **Table 3**. The first pair of stance columns uses micro-level position, and the second pair uses macro-level position. With only two retained stance categories, this specification is a binary logit contrast; the two reported marginal effects are equal in magnitude and opposite in sign.

* $p < .05$, ** $p < .01$, *** $p < .001$.

## 3.3 Downstream Outcome Profiles of Patent-Citing Papers

**Table 6** reports weighted post-matching regressions for the main 1:3 matched sample. The 1:3 specification is used as the main analysis, while estimates based on the alternative 1:1 and 1:5 matched samples are reported as sensitivity analyses in **Supplementary Table S8**. Each of the three patent-citation measures was entered in a separate specification.

**Table 6. Weighted Post-Matching Regressions of Downstream Paper-Level Outcomes**

| Outcome variable | Patent−citing indicator | ln(Patent citation count+1) | IHS(Patent citation ratio) | Observations |
|---|---|---|---|---|
| ln(FWCI+1) | −0.092***<br>(0.023) | | | 2625 |
| | | −0.081***<br>(0.021) | | |
| | | | −1.259<br>(0.873) | |
| ln(Cited−by count+1) | −0.129***<br>(0.032) | | | 2632 |
| | | −0.097**<br>(0.033) | | |
| | | | −1.176<br>(1.352) | |
| ln(Patent count+1) | 0.033**<br>(0.012) | | | 2632 |
| | | 0.021<br>(0.011) | | |
| | | | 0.301<br>(0.270) | |
| ln(Newsfeed count+1) | 0.002<br>(0.006) | | | 2632 |
| | | −0.002<br>(0.004) | | |
| | | | −0.045<br>(0.058) | |
| ln(NCT count+1) | −0.002*<br>(0.001) | | | 2632 |
| | | −0.002***<br>(0.001) | | |
| | | | −0.019<br>(0.012) | |
| IHS(Disruption) | −0.001**<br>(0.000) | | | 2632 |
| | | −0.000 | | |
| | | (0.000) | | |
| | | | −0.007 | |
| | | | (0.007) | |
| Retraction indicator | −0.001 | | | 2632 |
| | (0.002) | | | |
| | | 0.001 | | |
| | | (0.003) | | |
| | | | 0.024 | |
| | | | (0.052) | |
| Controls | Yes | Yes | Yes | |
| Year Fixed Effects | Yes | Yes | Yes | |
| Article-Type Fixed Effects | Yes | Yes | Yes | |
| Domain Fixed Effects | Yes | Yes | Yes | |

Notes: Coefficients from weighted post-matching regressions are reported, with cluster-robust standard errors in parentheses. Each regression includes only one key explanatory variable. For each outcome, estimates from three separate regressions are grouped in consecutive rows; the three patent-citation measures are not entered simultaneously. IHS denotes the inverse hyperbolic sine transformation.
* p < .05, ** p < .01, *** p < .001.

In the main 1:3 specification, the patent-citing indicator was negatively associated with scholarly-citation outcomes. The indicator was associated with lower ln(FWCI+1) (coefficient = −0.092, p < .001) and ln(total cited-by count+1) (coefficient = −0.129, p < .001). It was also negatively associated with the clinical-trial-reference measure (coefficient = −0.002, p < .05) and disruption (coefficient = −0.001, p < .01) in this specification.

The association with subsequent patent citations differed in direction: the patent-citing indicator was positively associated with citations received from later patents (coefficient = 0.033, p < .01). Its associations with news mentions and recorded retraction status were not statistically significant.

The patent-citation-count specification showed a similar pattern for scholarly-citation outcomes. The count measure was negatively associated with FWCI (coefficient = −0.081, p < .001) and total cited-by count (coefficient = −0.097, p < .01). It was also negatively associated with clinical-trial references (coefficient = −0.002, p < .001). Its associations with subsequent patent citations, news mentions, disruption, and recorded retraction status were not statistically significant.

The patent citation ratio was not statistically significantly associated with any examined outcome in the reported specifications. The pattern of statistical significance therefore varied across measures, but significance for one measure and nonsignificance for another does not itself establish that their associations differ. The measures use different scales and transformations, and the null results do not rule out associations that the present data estimate imprecisely.

The negative associations with scholarly-citation outcomes and the positive association between the patent-citing indicator and subsequent patent citations were consistent across the 1:1, 1:3, and 1:5 matching specifications. By contrast, the disruption estimate and the association between the patent-citing indicator and clinical-trial references were not stable across matching ratios. The patent-citation ratio remained statistically nonsignificant in all three matched samples (Supplementary Table S8).

## 4. Discussion and Conclusion

This study examined paper-to-patent citation through three connected dimensions: the contexts in which patents are cited, the associations among represented content, citation motivation, textual position, and evaluative stance, and the downstream outcome profiles associated with patent citation. The descriptive results first establish the empirical baseline against which the multidimensional analysis should be interpreted. Explicit patent citation was rare in the observed AI-related PMC corpus, consistent with earlier evidence that scientific papers only infrequently cite patents (Hsiao & Torvik, 2019). Most identified citations were also neutral and background-oriented, broadly consistent with earlier full-text evidence that patent citations often occur near the beginning of scientific articles and are frequently mentioned only once (Hsiao & Torvik, 2020). The contribution of these results lies less in establishing these descriptive regularities than in showing why they are insufficient for interpreting paper-to-patent citation as a uniform bibliographic signal. A patent reference is a selective trace whose substantive meaning depends on how the patent is used in the citing text, consistent with broader warnings that citation-based indicators provide incomplete evidence of knowledge flows (Meyer, 2000; Roach & Cohen, 2013).

The principal contribution of the citation-level analysis is the integrated multidimensional framework for distinguishing different forms of engagement with patented technology. By jointly considering represented content, citation motivation, textual position, and evaluative stance, the framework differentiates bibliographically similar patent citations that may reflect substantively different relations between scientific research and patented technology. Citation research has long emphasized that references can perform different rhetorical and epistemic functions within scientific argument (Gilbert, 1977; Small, 1978), while full-text studies have shown that citation function is systematically related to textual position and article structure (Boyack et al., 2018; Cohan et al., 2019; Teufel et al., 2006). The present findings extend this perspective to patent documents by showing systematic relationships among what technological content is represented, why the patent is cited, where the citation appears, and how the patented technology is evaluated. Knowledge-boundary perspectives provide a useful interpretive vocabulary for these patterns. Patent documents can retain their identity while being selectively represented for different purposes across technological and scientific communities (Star & Griesemer, 1989), while Carlile's (2004) distinction among transfer, translation, and transformation helps distinguish relatively shallow representation from deeper operational engagement. These concepts provide interpretive analogies rather than directly measured processes in this study. Background citations are consistent with contextual representation, whereas adoption, comparison, extension, or criticism can indicate more explicit textual engagement. Reported adoption is not verified implementation, and criticism or comparison does not by itself demonstrate technological uptake.

The pairwise and conditional associations suggest three heuristic configurations. The first is contextualization: background-oriented citations tend to invoke application scenarios or system architectures, appear in introductory or prior-work sections, and adopt neutral stances. In this configuration, the patent primarily situates the research within an existing technological environment. The second is operationalization: adoption-oriented citations are more likely to involve algorithms, models, devices, or workflows, occur in methods-related sections, and express supportive evaluations. Here, the citing text presents the patent as a technical or procedural resource; the analysis does not independently verify implementation. The third is problematization: patents cited to establish gaps or limitations tend to occur in introductory, prior-work, or evaluative contexts and to receive critical or mixed assessments. These configurations are heuristic summaries of observed associations rather than latent classes or causal mechanisms. Their significance lies in showing that identical bibliographic links can encode qualitatively different forms of engagement with patented technology. Their potential value is methodological: they identify ambiguities that reference counts alone cannot resolve. Establishing the frequency and reliability of the full configurations requires joint classification counts and independent validation beyond the pairwise patterns reported here.

The downstream analysis provides a complementary paper-level perspective. Patent-citing papers did not constitute a uniformly higher- or lower-impact class of research. After matching and covariate adjustment, the presence of a patent citation was associated with lower scholarly-citation measures but greater visibility to subsequent patents. The pattern of statistical significance varied across the binary, count, and share specifications. This variation does not establish differences in informational content, because the measures differ in scale, transformation, and estimation precision. In particular, these models do not identify a dose–response relationship between patent-citation intensity and technology-to-science linkage. The positive association with patent citations is consistent with a technological-orientation interpretation, but does not validate that interpretation. The negative scholarly-citation associations describe a different outcome pattern within the matched sample. These associations should not be interpreted as causal effects or as direct measures of research quality. One possible interpretation of this differentiated profile is that scholarly and technological visibility reflect partly different orientations and audiences. Scientific and technological knowledge production place different relative weights on theoretical novelty, methodological credibility, public disclosure, implementation, usefulness, and appropriability (Partha & David, 1994; Sauermann & Stephan, 2013). These orientations overlap, and university–industry engagement does not necessarily diminish academic performance (Perkmann et al., 2013), but research involving devices, deployment environments, engineering constraints, workflows, or mature technological systems may attract technological attention without becoming equally central to broad scholarly citation networks. The present analysis does not test the proposed orientation-and-audience explanation. It also does not observe patent ownership, legal status, licensing, enforcement, or freedom-to-operate conditions (Murray & Stern, 2007). The results therefore describe associations with different forms of recorded visibility, rather than a demonstrated mechanism linking citation context to later recognition.

The main practical implication concerns the use of paper–patent links in bibliometric, science-of-science, and research-evaluation studies. Patent references should not be treated as interchangeable evidence of technology-to-science knowledge transfer. A background mention, operational adoption, comparison, deployment example, or criticism represents a qualitatively different relation between the citing paper and the patented technology. Reference-list counts alone cannot distinguish these relations. Likewise, the number or ratio of patent citations should not automatically be interpreted as the strength of technological dependence or practical impact. Where paper-to-patent citations are used as indicators, incorporating citation function, represented content, textual position, or other full-text information can provide a more defensible interpretation of what the link captures. For research evaluation, the differentiated scholarly and technological outcome profiles also caution against relying on academic citation indicators alone when technological visibility is relevant. At the level of scholarly reporting, traceable patent identifiers and clear descriptions of how a patent is used can further improve the interpretability of these links.

Overall, this study clarifies both the informational content and the interpretive limits of paper-to-patent citation. In the AI-related PMC corpus, explicit patent citations were uncommon and predominantly contextual, while a smaller subset entered the methodological or evaluative core of papers through adoption, comparison, modification, deployment, or criticism. The heuristic configurations of contextualization, operationalization, and problematization illustrate that the same bibliographic relation can represent substantially different forms of engagement with patented technology. At the paper level, the patent-citing indicator was associated with different scholarly and technological outcome profiles. Differences among the binary, count, and share specifications cannot be established from their significance patterns alone. Paper-to-patent citation should therefore be interpreted as a selective and heterogeneous bibliographic signal of contact between scientific and technological discourse rather than as a direct or general indicator of technology-to-science knowledge transfer, research quality, or research value.

Several limitations should be noted. First, the study is based on PMC full text and a keyword-defined AI corpus, which may omit relevant AI studies, include papers in which AI is only peripheral, and underrepresent computer science conference proceedings, engineering literature, industrial reports, and proprietary publisher content. Although the rarity of patent citation is broadly consistent with previous large-scale studies of paper-to-patent citation (Hsiao & Torvik, 2019, 2020), its absolute prevalence should be interpreted cautiously because reliance on structured JATS reference metadata may lead to the underidentification of untagged, malformed, or nonstandard patent references. Accordingly, the non-patent-citing controls are XML-screened rather than independently verified through a separate patent-number-based procedure. Second, the citation-context classifications remain model-dependent despite the use of three large language models, conservative context construction, majority voting, textual evidence, and confidence scores. Without an expert-annotated benchmark, some estimated associations may reflect prompt design, overlapping category definitions, or correlated classification errors. Third, exclusion of rare categories restricts the regression results to the retained classes, and repeated instances within papers limit the number of independent observations. The association models and matching analyses do not identify causal mechanisms or effects, and unobserved differences may remain between patent-citing and non-patent-citing papers. Interpretation of the recorded outcomes also depends on their temporal coverage and the available follow-up period. Finally, the analysis is limited mainly to the paper side of paper–patent relations. It does not examine the cited patents' full texts, claims, legal status, ownership, value, licensing conditions, or citation trajectories, and the selected outcomes do not directly capture reproducibility, commercial success, clinical effectiveness, policy use, or broader social benefit. Future research could extend the framework to broader publication venues and technical fields, integrate paper-side citation contexts with patent-side textual and legal attributes, develop expert-validated annotation benchmarks, and examine whether different forms of patent citation correspond to distinct long-term pathways of scientific, technological, clinical, and social visibility.

## Author Contributions

Haowei Li: Conceptualization; data curation; formal analysis; investigation; methodology; software; visualization; writing—original draft; writing—review and editing.

Yi Bu: Conceptualization; funding acquisition; project administration; supervision; writing—review and editing.

## Acknowledgements

This work was supported by the National Social Science Foundation of China (#24&ZD072). The authors are grateful to all members at the Knowledge Discovery Lab, Peking University.

## Conflict of Interest Statement

The authors declare no conflicts of interest.

## Data Availability Statement

The data and code supporting the findings of this study will be made publicly available upon acceptance of the manuscript at https://github.com/haoweili02/AI_research_papers_citing_patents.

## Generative AI Statement

During the preparation of this manuscript, the authors used DeepSeek-V3.2-Reasoner, Gemini-3.1-Flash-Lite-Preview, and GPT-5.4-mini in combination to assess the relevance of candidate citation-context sentences and to classify citation contexts according to the predefined framework. These models were used as computational annotation tools. The authors also used Codex to assist with manuscript review, including checks of argumentation, methodological reporting, consistency between tables and text, and language revision. These review suggestions do not constitute new empirical analyses or independent validation of the data. The authors remain responsible for verifying all AI-assisted suggestions and for the final methodological choices, interpretation, conclusions, and manuscript content.